\documentclass{knawproc}
\usepackage{psfig}

\begin{document}

\begin{opening}

\title{Radio Observations of High Redshift \\ Radio Galaxies}

\author{C.L. Carilli$^1$ H.J.A. R\"ottgering$^2$ G.K. Miley$^2$ \\
L.H. Pentericci$^2$ D.E. Harris$^3$} 
\addresses{
~$^1$National Radio Astronomy Observatory, Socorro, NM, USA \\
~$^2$Leiden Observatory, Leiden, The Netherlands \\
~$^3$Smithsonian Astrophysical Observatory, Cambridge, MA, USA \\
}

\runningtitle{Radio Observations}
\runningauthor{Carilli et al.}

\end{opening}

To appear in: {\sl The Most Distant Radio Galaxies},
eds. H.J.A. R\"ottgering, P.N. Best, and M.D. Lehnert (North-Holland:
Amsterdam). 


\begin{abstract}

We review some aspects of radio continuum polarimetric imaging of high
redshift radio galaxies. The correlation between extreme values of
Faraday rotation observed toward radio emitting structures in nearby 
radio galaxies, and X-ray emitting cluster atmospheres, is presented 
as a method for  targeting objects at high redshift for deep X-ray
searches. We present an X-ray detection of the extreme rotation
measure radio galaxy PKS 1138$-$262 at $z = 2.156$, and we argue that
the X-ray emission is from a cluster 
atmosphere with a luminosity of 1.7$\pm$0.3$\times$10$^{44}$
h$^{-2}$ ergs  sec$^{-1}$. 
We also present results on the correlation between size and
redshift for a sample of ultra-luminous radio galaxies between 0 $<$ z
$<$ 4.3. Source sizes decrease systematically  with redshift, suggesting
either denser environments, or younger sources, at high redshift. An
alternative explanation is significant inverse Compton losses off the
microwave background at high redshift. 

\end{abstract}


\section{Introduction}

Searches for high redshift radio galaxies begin with 
surveys of the sky at radio frequencies typically below 1.4 GHz. 
Although these sources are  targeted for study due to their
radio properties, subsequent work has focused primarily on the optical
and near IR properties of these systems. Sensitive, high resolution, 
multifrequency radio continuum polarimetric imaging provides
unique information about the physics of these sources, and their
environments, in a number of ways, including: 
(i) identification of the location of the active nucleus, 
(ii) study of alignments between radio and optical structures on
sub-kpc scales (Best et al. this volume, van Breugel et al. this volume),
(iii) determination of the pressure in the ambient medium, 
(iv) determination of the strength and morphology of source magnetic
fields, (v) searching for small scale gravitational lensing events 
towards the extended radio structures 
(Carilli, Owen, and Harris 1994, Carilli 1995, Kochanek and Lawrence 1990),
(vi) searching for extreme rotation measures and ultra-steep spectrum regions
(Carilli et al. 1997b), 
(vii) study of the cosmological evolution of radio source structure
(Barthel and Miley 1989, Singal 1993), 
(viii) tests of quasar-radio galaxy unification schemes as a function of 
redshift (Antonucci 1993), and 
(ix) constraining basic cosmological parameters (Guerra
and Daly 1998).

An extensive review of jet theory for  powering the double radio lobes
in powerful extragalactic radio sources can be found in Carilli and
Barthel (1996).
In this paper we consider a few topics of particular interest for
understanding the environment and evolution of 
powerful radio galaxies at high redshift. 

\section{Extreme Rotation Measures: Cluster Atmospheres
at High Redshift}

Rich clusters of galaxies constitute the extreme high
mass end of the cosmic mass distribution function, and  study of such rare
objects provides unique leverage into theories  of structure
formation in the universe (Peebles 1993). 
Unfortunately, optical studies of distant clusters can 
be corrupted by field galaxy contamination (Frenk et al. 1990), while
searching for X-ray emission from cluster atmospheres at z $>$ 1
requires very long integration times for marginal
detections. Thus far the  three most distant X-ray cluster candidates
currently known are all associated with powerful radio sources at z
$\approx$ 1 (Crawford and Fabian 1993, Crawford and Fabian 1995, Dickinson  
1997, Worrall et al. 1994). 

One of the more important results in the study of extragalactic radio
sources  in the past 10 years has been the discovery of extreme values
of Faraday rotation toward extended radio emitting structures (Dreher,
Carilli, and Perley 1987, Taylor, Barton, and Ge 1994, Carilli et
al. 1997a). Rotation measure (RM) values up to 20000 rad m$^{-2}$ have
been detected, as well as gradients in RM $\ge$ 1000 rad
m$^{-2}$ arcsec$^{-1}$. Observations
of lower redshift radio galaxies have shown a  clear correlation between 
extreme rotation measures and cluster environment: all sources 
located at the centers of dense, X-ray emitting cluster atmospheres show
large amounts of Faraday rotation (Taylor et al. 1994). 
An important point is that this correlation is  independent
of radio source luminosity and morphological class, 
and hence is most likely a  probe of cluster
properties, and not radio source properties. The implication is that
the hot cluster gas must be substantially magnetized, with field strengths of
order a few $\mu$G (Carilli et al. 1997a). 

The correlation between cluster X-ray properties and extreme rotation
measures provides a potentially powerful criterion for selecting high z
sources for deep  X-ray observations. 
We have been conducting a
systematic search for extreme RM sources at high redshift in order to
exploit this correlation, and thereby potentially discover X-ray
emitting clusters at high redshift (Carilli et al. 1997b). 

Our best example to date is the narrow emission line radio galaxy PKS
1138$-$262 at z = 2.156. This source has  the largest known RM at z
$>$ 2, with (rest frame) RM values up to 6250 rad 
m$^{-2}$ (Carilli et al. 1997b).
The optical continuum morphology of 1138$-$262 is comprised of many
blue `knots' of emission, distributed over an area of  about 100 kpc
(Figure 1). The Ly$\alpha$ emission shows similar
clumpy structure on a similar scale.
Pentericci et al. (1997) interpret this distribution as a group of
star forming galaxies in the act of merging into a giant elliptical
galaxy. The radio source has the most disturbed morphology
of any z $>$ 2 radio galaxy yet identified,  indicating strong
interaction between the radio jet and the ambient medium.
Such interaction is supported by optical
spectroscopy, showing large velocities, and velocity dispersions, for
the line emitting gas in regions associated with the brightest radio
knots. All these data provide strong, although circumstantial,
evidence that the 1138$-$262 radio source is in a dense environment,
perhaps a hot, cluster-type atmosphere. 
To test this hypothesis we observed 1138$-$262 with the High
Resolution Imager on the ROSAT X-ray satellite (Carilli et al. 1998).

\begin{figure}
\psfig{figure=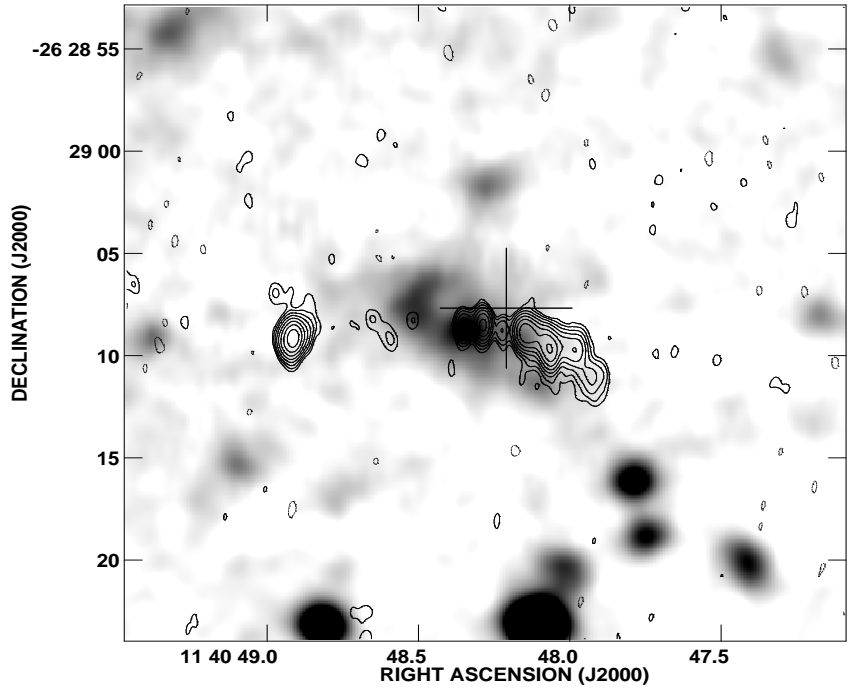,width=4.5in,height=4.5in}
\caption{The 
radio image of 1138$-$262 at 5 GHz (contours) with 
a resolution of 0.4$''$$\times$0.7$''$. 
The contour levels are a geometric
progression in square root two, with the first level being 
0.15 mJy beam$^{-1}$.
The grey-scale shows the optical image (R + B bands) of 1138$-$262. 
The cross indicates the position of X-ray peak surface brightness, and 
the cross size indicates the estimated astrometric accuracy.
}
\end{figure}

\begin{figure}
\psfig{figure=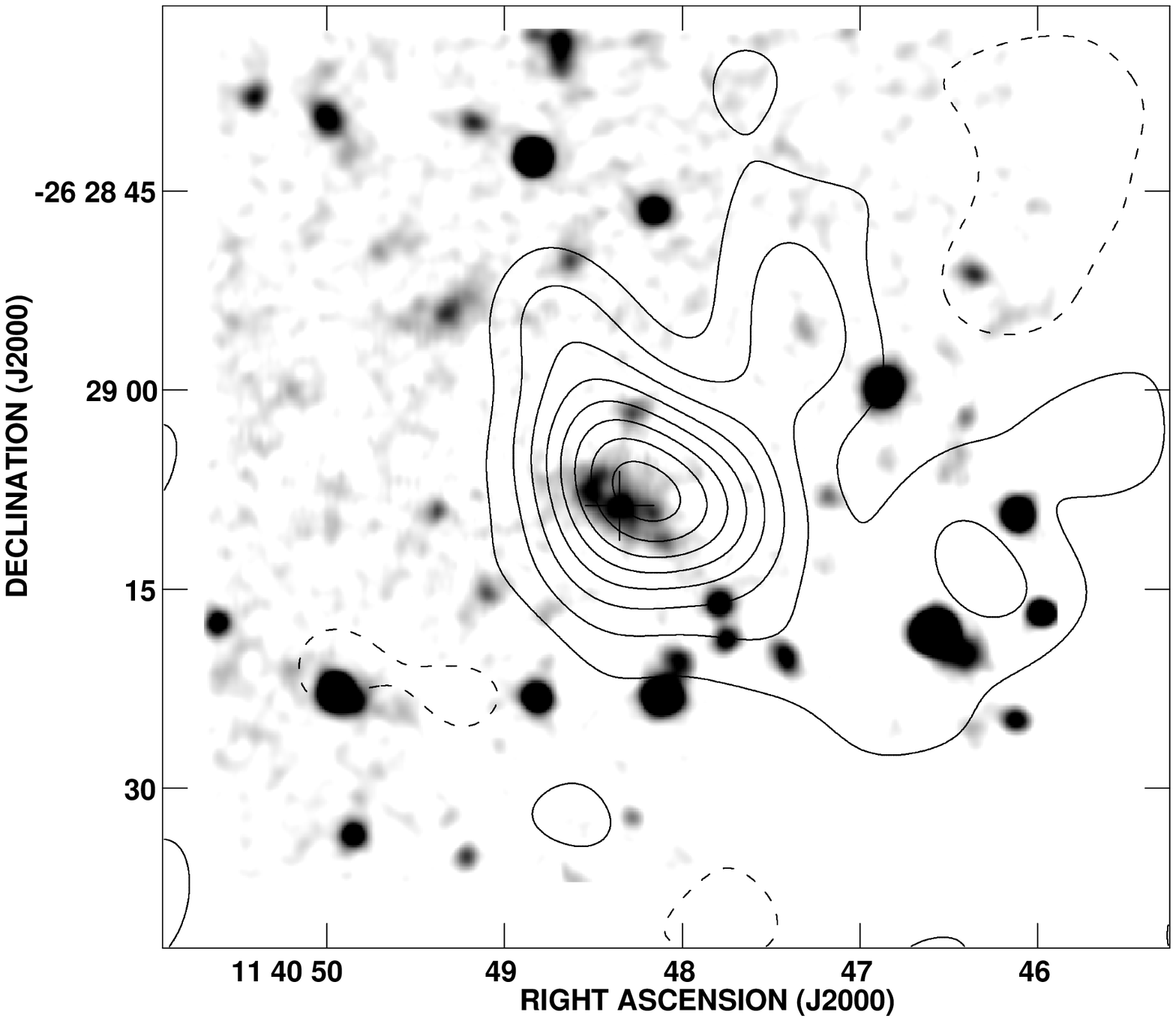,width=4.5in,height=4.5in}
\caption{The X-ray image of   1138$-$262 from a
41 ksec exposure with the ROSAT HRI
(from Carilli et al. 1998).
The image has been convolved with a Gaussian beam of FWHM = 10$''$,
and a   mean background level of 5 counts per beam has been subtracted. 
The contour levels are: -4, -2, 2, 4, 6, 8, 10, 12, 14, 16 counts per beam.
The cross shows the position of the radio and optical nucleus
of 1138$-$262.
}
\end{figure}

The ROSAT observation shows a clear
detection of an X-ray source close to the position of the radio and 
optical nucleus of 1138$-$262 (Figure 2).
The X-ray luminosity for 1138$-$262 is
1.7$\pm$0.3$\times$10$^{44}$ h$^{-2}$ ergs 
sec$^{-1}$ for emitted energies between 2 keV and 10 keV.
The optical, radio, and X-ray data all favor 
a  hot cluster gas origin for the X-rays from
1138$-$262, however we cannot preclude an AGN contribution 
(see Carilli et al. 1998 for a detailed discussion).
If the X-ray emission is from a hot cluster atmosphere, then the
total gravitational mass is a few$\times$10$^{14}$
M$_\odot$, and the total gas mass is a few$\times$10$^{13}$
M$_\odot$. 

We proceed under the assumption that the X-ray emission from 
1138$-$262 is  from a typical massive cluster atmosphere at z = 2.156,
and we speculate on possible  physical implications. First, 
models for structure formation such as CDM using
standard cosmologies predict that 
any large scale structure observed at z
$\approx$ 2 must be dynamically young, having just separated from the 
general expansion of the universe (Peebles 1993). Alternatively,
massive structures 
at large redshift might indicate  an open universe ($\Omega$ $<<$ 1).
Second,  the existence of a hot cluster-type atmosphere at high redshift
would argue for  shock heated,  primordial in-fall as the origin of
the cluster gas, rather than ejecta from cluster galaxies (Sarazin 1986). 
Such a model would require an early
epoch of star formation 
in order to supply the metals observed in the associated optical
nebulosity (Ostriker and Gnedin 1996). Third, the large rotation
measures observed  raise the question of the origin of  cluster
magnetic fields at early epochs. One possible mechanism is  
amplification of seed fields by a turbulent dynamo  operating 
during the formation process of the cluster as driven
by successive accretion events (Loeb and Mao 1994). And forth, 
the observation of X-ray emission from 1138$-$262  would then  
verify the technique of using extreme rotation
measure sources as targets for searches for cluster X-ray emission at high
redshift. Thus far there have been a total of 11 radio galaxies at
z $>$ 2 with source frame rotation measures greater than 1000 rad
m$^{-2}$, the most distant being  at z = 3.8 (Carilli et al. 1997b,
Ramana et al. 1998, Carilli, Owen, and Harris 1994).

An interesting alternative to a massive cluster atmosphere 
giving rise to the X-ray emission from 1138$-$262 is thermal
emission from a very dense, sub-cluster `halo', perhaps associated
with the (forming) cD galaxy on 
a scale $\le$ 100 kpc. For  instance, increasing the gas density from
0.01 cm$^{-3}$ to 0.1 cm$^{-3}$  would decrease the required hot gas
mass by the same factor (at fixed X-ray luminosity), and could
possibly alleviate constraints on 
cosmological structure formation models. The pressure in this gas
would be very high (10$^{-9}$ dynes cm$^{-2}$), comparable to the
pressure in the optical line emitting  
nebulosity and to the minimum pressure in the radio source, and the
cooling time would be short ($\le$ few$\times$10$^{8}$ years).
Circumstantial evidence for such very dense, hot gas enveloping some
high z radio sources has been reviewed by Fabian et al. (1986). Fabian
(1991) suggests that in some cases the `inferred pressures are close
to the maximum that can be obtained by gas cooling in a potential well
of a galaxy' (ie. the cooling time $\approx$ gravitational free-fall
time), and he designates such systems as  `maximal cooling
flows', with implied cooling flow rates up to 2000 M$_\odot$
year$^{-1}$.  High resolution  X-ray imaging with AXAF should be able
to test whether 1138$-$262  has a normal cluster atmosphere, a `maximal
cooling flow', or an unusually X-ray loud AGN. 
  
\section{Properties of  Sources at High Redshift}
\subsection{The Ultra-luminous Sample}

An extensively investigated property of extragalactic radio sources has
been the relationships between radio power (P), source size (D), 
and redshift (z),   i.e. the
P-D-z relationships (Singal 1993, Blundell et al. 1996). 
Such studies address the possibility of using radio sources as 
standard-rulers for cosmological models ({\sl cf.}
Daly 1994), or perhaps more reasonably as probes of the evolution
of the local gaseous environments of the sources ({\sl cf.} Neeser  et al.
1995, Eilek and Shore 1989, Subrahmanian and Swarup 1990, 
Gopal-Krishna and Wiita 1991, Wellman and Daly 1995).

Although much work has been done on this topic, there remains fundamental
disagreements between various studies. Some studies favor a
steep evolution for source size with redshift: D $\propto$ (1 + z)$^{-n}$,
with $n$ $\ge$ 3, and favor a finite correlation between source
size and radio power: D $\propto$ P$^m$, with $m$ $\approx$
0.3$\pm$0.1 (Singal 1993, McCarthy this volume).
Other studies find a more gradual size evolution with redshift,
with $n$ $\approx$ 1.5, and find essentially no correlation between
power and size, $m$ $\approx$ 0 (Neeser et al. 1995, Eales
this volume).  Neeser  et al. (1995) show that some of these
differences could arise due to limitations in the data, including:
(i) the strong Malmquist bias inherent in  flux limited samples,  
(ii) the incompleteness
of samples, in particular due to surface brightness selection effects
and/or the lack of  redshifts for sources in a given
sample, (iii) mixing sources of different morphological class, and
including or excluding compact steep spectrum objects from a sample,
and (iv) limited redshift ranges covered for most samples.

The Leiden observatory has, for some years, been performing a 
systematic search for radio galaxies at high redshift under the
direction of G. Miley. We have assembled a data set of physical
parameters for a sample of ultra-luminous (`UL') radio
galaxies spanning the redshift range 0 to 4.3. The data for the high
redshift sources come mostly from the Leiden program, including data
from a follow-up high resolution, multifrequency radio continuum 
imaging survey of high redshift 
radio galaxies (Carilli et al. 1997b). These data have been supplemented
with data on lower redshift sources
from  the 3CR sample as defined and
compiled for FRII radio galaxies by McCarthy, van Breugel, and Kapahi (1991b), 
and data from other published samples of radio galaxies, including 
the Molongolo sample (McCarthy  et al. 1990, 1991a),
the B2 sample  (Owen and Keel 1995, Lilly 1988), 
and the 8C sample (Lacy et al. 1994). 

Figure 3 shows the distribution of rest frame spectral luminosity 
at 178MHz, P$_{178}$, versus redshift, 
for all 3CR Fanaroff-Riley class II radio galaxies, supplemented with 
sources from studies as listed above. Note the strong
correlation between P$_{178}$ and redshift in the 3C sample, and how
adding lower flux density sources beyond z~=~1 to the 3C sample `fills-in'
the distribution. The `Ultra-Luminous (UL) sample'  is defined 
as FRII radio galaxies with:~ 34.7 $\le$ log(P$_{178}$) $\le$ 36.3, ~
 i.e. sources with luminosities roughly within a factor of $\pm$ five of
Cygnus A (about 100 sources total). It is clear from Figure 
3 that by adding the lower flux density sources at high redshift
to the higher luminosity sources in the
3C sample, this UL sample reduces the effect of  the Malmquist 
bias inherent in the  3C sample alone, although becoming necessarily
sparse at z $\le$ 0.4. The UL 
sample has good statistics to high redshift, is fully 
identified, including spectroscopic redshifts, and
contains  only FRII class radio  galaxies.   
Perhaps most importantly, the UL sample is relatively restricted in
radio luminosity, thereby mitigating problems
with residual correlations in P-D, although information on the
correlation of source properties with power is lost.
Readhead et al. (1996) have proposed a
classification scheme for extragalactic radio sources based on source
sizes. In the context of their scheme, the UL sample contains no
Compact Symmetric Objects, 12$\%$ Medium Symmetric Objects, and 88$\%$
Large Symmetric Objects.

\begin{figure}
\psfig{figure=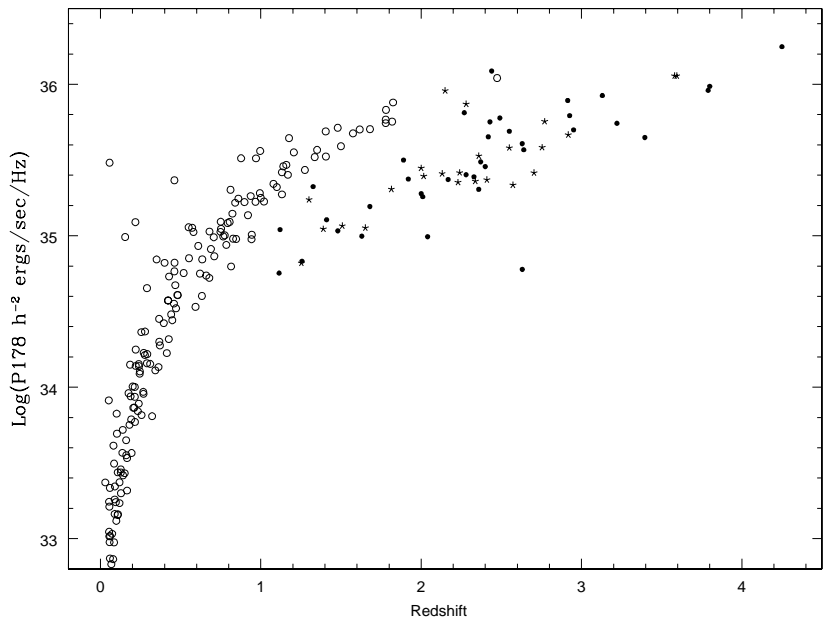,width=4.7in,height=3.5in,angle=-90}
\caption{A plot of rest frame spectral 
luminosity  at 178MHz (P$_{178}$)
versus redshift. The open circles are the Fanaroff-Riley class II
radio galaxies in the 3C sample. The filled circles are ultra-luminous
sources from other surveys without explicit maximum angular size cut-offs.
The stars are ultra-luminous sources from surveys with 
explicit maximum angular size cut-offs.
The `Ultra-Luminous (UL) sample'  is defined 
as FRII radio galaxies with:~ 34.7 $\le$ log(P$_{178}$) $\le$ 36.3.
}
\end{figure}

Many high z radio galaxy searches involve a radio spectral index cut-off
for source selection: $\alpha$ $\le$  -1.0 or so.  
The question arises as to 
what biases  such a selection introduces in the properties
of sources in the resultant sample.
We have used our database of integrated
spectra of radio galaxies to determine the redshift
evolution of radio spectral indices of the UL sample
at both a fixed observed-frame frequency, and a fixed source rest-frame
frequency. Figure 4a shows the spectral indices at an observed
frequency of 2 GHz for the UL sample.  Figure 4b shows
the spectral indices for the UL sample at a rest frame frequency of 2
GHz,  i.e. correcting for the redshift. Linear fits to the data
yield a slope for spectral index versus redshift of -0.12$\pm$0.05 for
the observed-frame data, and -0.08$\pm$0.05 for the K-corrected data,
where the errors imply unit changes in reduced $\chi^2$ assuming a
reasonable error of $\pm$0.1 for each spectral index measurement.
For the non-K-corrected data the median spectral indices
decrease from -0.94$\pm$0.08 for sources at z $<$ 1, to  -1.32$\pm$0.15
for sources at z $>$ 2.9, while for the K-corrected sources the
steepening  with redshift is 
reduced, going from -0.91$\pm$0.08 at z~$<$~1 to -1.08$\pm$0.15
at z $>$ 2.9, where the errors indicate the inner quartile range of the
distribution. Although the errors are significant, these data suggest
that the trend for spectral steepening with
increasing redshift  at a  fixed observing frequency is principally
due to the substantial  K-correction involved.
A similar conclusion was reached by van Breugel and McCarthy (1989).

\begin{figure}
\psfig{figure=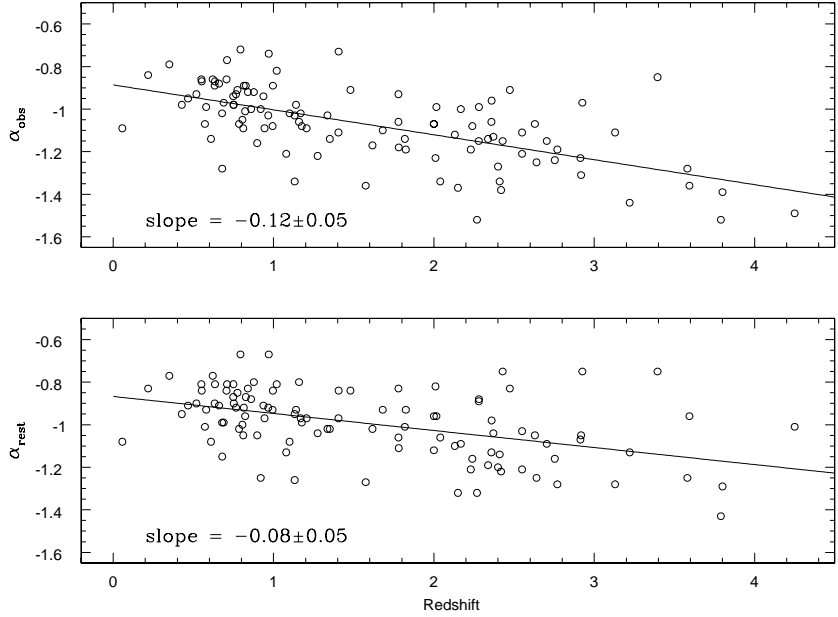,width=4.7in,height=3.5in,angle=-90}
\caption{The upper frame shows the spectral index versus redshift for the
ultra-luminous sample at an observed frequency of 2GHz.
The lower frame shows spectral index versus redshift for the UL sample
at a  source rest-frame frequency of 2GHz (i.e. K-corrected).
}
\end{figure}

Some of the searches  for high redshift radio galaxies include an
explicit angular size upper limit, typically $\le$ 20$''$.
For redshifts beyond 0.4 or so, this implies a physical size
limit of about 90 h$^{-1}$ kpc. In our analysis we keep track of sources
selected with and without explicit angular size cut-offs, and we find
no significant differences between the results. 
We also find that the 3C source size distribution joins smoothly to
that for the higher redshift sources. 

A more insidious problem is the angular size cut-off implicit
in identifying high redshift radio sources. An implicit cut-off might
arise from missing sources in a radio survey
due to the cosmological dimming in radio surface brightness. 
Again, selecting FRII radio sources, for which 
much of the emission comes from (typically) unresolved hot spots
at the source extremities, mitigates this problem. A more difficult 
problem is that of identifying the observationally faint high redshift
optical galaxies in large fields-of-view, as may occur in cases
of wide-separation double radio sources. 
Again, the 3C sample is completely
identified, as is the 6C sample of Neeser  et al.. Hence, the
`dove-tailing' of results for the high redshift sources in the 
UL sample with those for the  
3C sources, and the  agreement  between  results from 
the UL sample with those from 6C (see below), 
gives us hope that this latter effect is not dominating the statistics.

\subsection{\bf 4. The P-D-z-$\theta$ Relationships for the UL Sample}

Figure 5 shows  physical size versus redshift for the UL sample.
Physical sizes were derived from angular sizes
assuming a standard cosmology with q$_o$ = 1/2, 
and h $\equiv$ H$_o$/100 km sec$^{-1}$ Mpc$^{-1}$. 
The error bars show the range of  median size as
a function of redshift, for redshift bins chosen to have roughly
equal numbers of sources in each bin (about 20 sources per bin, except
for the highest redshift bin, which contains 12 sources).
The error bars indicate the inner quartile range
for the distribution within each bin. 
We have fit a power law in (1+z)
to the median sizes using three different weighting
schemes: weighted by the RMS for the distribution in each redshift bin,
weighted by the inner quartile range error bars, 
and an unweighted fit.
The results for the three different weighting methods are consistent,
within the errors, with:
$ D~=~170\pm40 h^{-1}(1 + z)^{-n}$ kpc, ~
with $n$~=~1.2 $\pm$ 0.3. The errors correspond to unit changes in
reduced  $\chi^2$ for a least squares fit weighted by the RMS in each
bin. We repeated the analysis using wider redshift bins, 
and using a narrower luminosity range,
and obtained essentially the same results.
We have also fit a power-law to all the data points (as opposed to the
median values), which yields: $ D~=~140\pm20 h^{-1} (1 + z)^{-n}$ kpc, ~
with $n$~=~1.2 $\pm$ 0.1. Again, the errors correspond to unit changes in
reduced  $\chi^2$ assuming a 10$\%$ error in each size measurement.

\begin{figure}
\psfig{figure=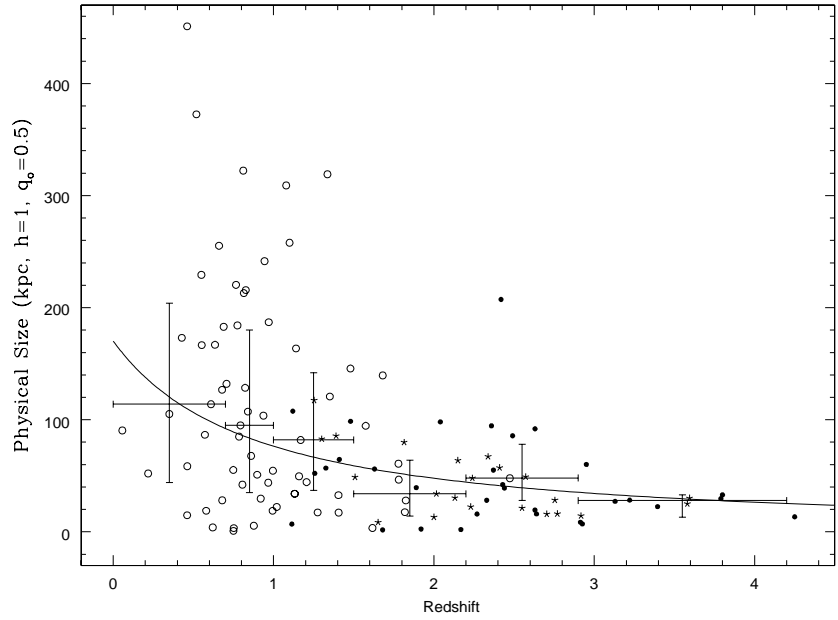,width=4.7in,height=3.5in,angle=-90}
\caption{A plot of physical size versus redshift for the UL sample.
symbols are the same as Figure 3.
The error bars indicate inner quartile regions for physical size in
various redshift bins, and the solid line is 
a power-law fit to the data, of the form:~ 
D~=~170$\pm$40h$^{-1}$(1 + z)$^{-n}$ kpc, with $n$~=~1.2 $\pm$ 0.3.
}
\end{figure}

We have also  considered the evolution of
source bending angle, $\theta$,  as a function
of redshift, where $\theta$ is defined as 180$^o$ minus 
the angle between the two lines joining the core with the hot spot
on either side of the source.
Figure 6 shows the relationship between $\theta$ and redshift for the UL
sample for sources with clearly defined hotspots and cores.
The general trend is for increasing bending angle with redshift,
at least out to z $\approx$ 3.
The median value behaves as: $\theta$ = $6^o \pm 3^o$ for z $<$ 1, 
$\theta$ = $10^o \pm 4^o$ for 1 $<$ z $<$ 2,
$\theta$ = $14^o \pm 3^o$ for 2 $<$ z $<$ 2.9, and
$\theta$ = $10^o \pm 5^o$ for 2.9 $<$ z.
A similar increase in bending angle with redshift has been seen by
Barthel and Miley (1988) for radio quasars. The angles involved for the
quasars are a factor two or so larger than those for radio galaxies --
qualitatively consistent with unification schemes, i.e. larger
projection angles for quasars relative to radio galaxies.

\begin{figure}
\psfig{figure=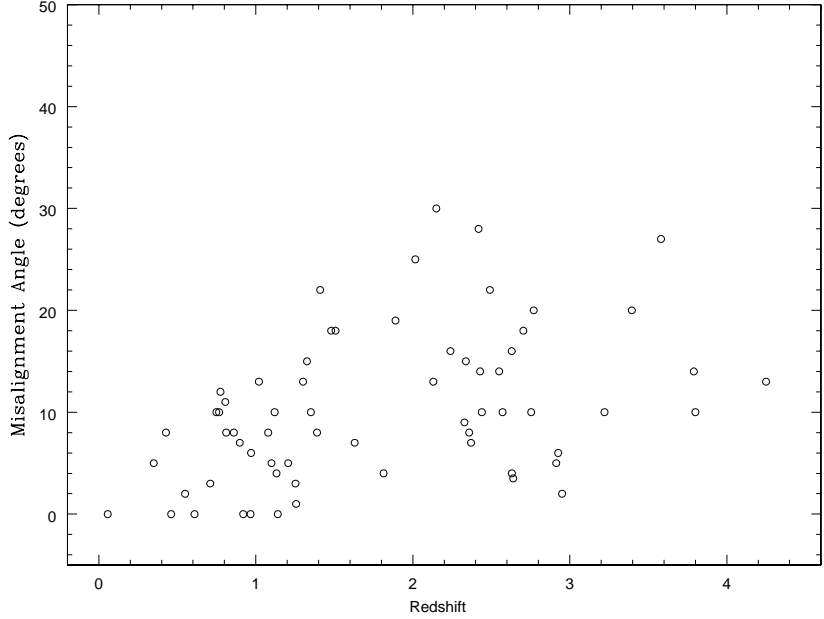,width=4.7in,height=3.5in,angle=-90}
\caption{A plot of radio source bending angle versus redshift for the UL
sample.  The bending angle, $\theta$, is 
defined as 180$^o$ minus 
the angle between the two lines joining the core with the hot spot
on either side of the source.
}
\end{figure}

One possible explanation for the decrease in source size with redshift
at `fixed' radio luminosity is an increase in the external density
with redshift. Barthel and Miley (1988) show that the observed value
of $n$ $\approx$ 1.5 is consistent with a simple model in which 
the co-moving gas density evolves 
cosmologically: n$_x$ $\propto$ (1 + z)$^3$. This leads to 
source size evolution according to: D $\propto$ (1 + z)$^{-1.5}$,
assuming the ram pressure advance speed for a typical source 
behaves as n$_x^{-0.5}$.
Denser environments at high redshift could also explain the increase
in bending angle with redshift, assuming that source bends are due to
jet-cloud collisions (Icke 1991).
A possible problem with the simple model of Barthel and Miley is
that that the dynamical models of Begelman (1996) and Readhead et
al. (1996) suggest that the source advance speed may have a weaker
dependence on external density than n$_x^{-0.5}$. Also, it is not clear
that the local density of a structure that has separated from the
general Hubble expansion will evolve as the cosmic
background  density. Daly (1998) has presented another model in which 
the decrease in source size with redshift is due to the sources being
systematically younger at high redshift.

Overall, the data indicate that sizes of ultraluminous 
radio sources decrease with redshift. However, there remains
a fundamental, unexplained, discrepancy between various studies 
of the D-z relationship, with some studies favoring a fairly
shallow evolution, with an index of  n
$\approx$ $-1.5$ (the UL sample, and the study
of Eales et al. this volume) , and others studies
favoring a steep evolution, with an index of n $\approx$ $-3.0$
(McCarthy et al. this volume). 
The cosmic evolution of physical sizes of radio sources is
a potentially powerful probe of the evolution of
source physics, and/or  source environments, and hence
is an important issue to address with deeper, complete samples of
radio sources now becoming available. 

\subsection{Inverse Compton Losses}

An interesting alternative explanation of the D-z relationship for
ultraluminous radio galaxies 
is that the larger sources may be `snuffed-out' by inverse
Compton losses at some nominal size which depends on redshift.
The relative importance of inverse Compton losses versus synchrotron
losses for the relativistic electrons in radio sources is hypothesized
to be a strong function of redshift (Krolik and Chen 1991, Daly 1992).
The `characteristic frequency', $\nu_*$,
above which one should detect exponential 
spectral steepening due to inverse Compton 
losses behaves as:~ $\nu_{*}$ $\propto$ (1 + z)$^{-8}$ 
(measured in the source frame).
This steep redshift dependence of $\nu_*$
is due to the fact that the
energy density in the microwave background, U$_\gamma$, behaves as:~ 
U$_\gamma$ = 4.16x10$^{-13}$ (1 + z)$^4$ ergs cm$^{-3}$. At z = 3 
this corresponds to a magnetic field of strength 50$\mu$G, 
comparable to the equipartition fields found for 
the lobes of  ultraluminous radio galaxies.
The majority of the sources in the UL sample were selected 
from flux limited samples at frequencies between 178 MHz and 408 MHz,
implying rest frame frequencies of 800MHz to 1600 MHz at z = 3.
The implied synchrotron/inverse Compton lifetimes for the radiating
electrons are about 4 Myr. Adopting the  nominal source advance 
speed of 0.02 times the speed
of light proposed by Readhead et al. (1996) implies that the spectra
of the high 
z sources in the sample might be exponentially cut-off when they
reach radii of order 20 h$^{-1}$ kpc, a value consistent with 
the lack of sources larger than about 40 h$^{-1}$
kpc beyond z = 3 in the UL sample. 

An argument against this idea is the lack
of a strong residual redshift dependence in the K-corrected spectra
shown in Figure 4b. The counter argument is that, 
if the cut-off is exponential, the affected sources may 
simply drop-out of the flux limited samples. These sources would
appear as `naked hot spots and cores', i.e. only sites of very active
particle acceleration will be observed.

\begin{acknow}
The National Radio Astronomy Observatory is operated by Associated
Univ. Inc., under contract with the National Science Foundation.
We acknowledge support by a programme subsidy provided by the Dutch
Organization for Scientific Research (NWO). We thank J.P. Leahy for
important comments on portions of this paper.
\end{acknow}

\end{document}